\documentclass[conference]{IEEEtran}
\IEEEoverridecommandlockouts
\usepackage{cite}
\usepackage{amsmath,amssymb,amsfonts}
\usepackage{algorithmic}
\usepackage{graphicx}
\usepackage{textcomp}
\usepackage{xcolor}
\usepackage[hidelinks]{hyperref}

\newcommand{\expnumber}[2]{{#1}\mathrm{e}{#2}}

\usepackage{enumitem}
\newlist{questions}{enumerate}{2}
\setlist[questions,1]{label=RQ\arabic*.,ref=RQ\arabic*}
\setlist[questions,2]{label=(\alph*),ref=\thequestionsi(\alph*)}

\usepackage{mathtools}
\DeclarePairedDelimiter\ceil{\lceil}{\rceil}

\usepackage{multirow}
\usepackage{array}
\usepackage{subcaption}

\newcommand{\ie}{\textit{i.e.}, } 
\newcommand{\eg}{\textit{e.g.}, } 
\newcommand{\aka}{\textit{a.k.a. }}
\newcommand{\etc}{\textit{etc.}} 
\newcommand{\vs}{\textit{versus} } 
\newcommand{\etal}{\textit{et al.} } 
\newcommand{\naive}{na\"\i{}ve } 

\newcommand\blfootnote[1]{%
  \begingroup
  \renewcommand\thefootnote{}\footnote{#1}%
  \addtocounter{footnote}{-1}%
  \endgroup
}

\def\BibTeX{{\rm B\kern-.05em{\sc i\kern-.025em b}\kern-.08em
    T\kern-.1667em\lower.7ex\hbox{E}\kern-.125emX}}
\begin{document}

\title{Learning Compact Compositional Embeddings via Regularized Pruning for Recommendation
}



\author
{
 Xurong Liang{\small$^\dag$}\hspace*{10pt}Tong Chen{\small$^\dag$}\hspace*{10pt}Quoc Viet Hung Nguyen{\small$^\perp$}\hspace*{10pt}Jianxin Li{\small$^\S$}\hspace*{10pt}Hongzhi Yin{\small$^{\dag *}$}\\
 \fontsize{10}{10}\selectfont\itshape $~^\dag$The University of Queensland, Australia, {\fontsize{9}{9}\selectfont\ttfamily\upshape \{xurong.liang,tong.chen,h.yin1\}@uq.edu.au}\\
  \fontsize{10}{10}\selectfont\itshape $~^\perp$Griffith University, Australia, {\fontsize{9}{9}\selectfont\ttfamily\upshape henry.nguyen@griffith.edu.au}\\
   \fontsize{10}{10}\selectfont\itshape $^\S$Deakin University, Australia, {\fontsize{9}{9}\selectfont\ttfamily\upshape jianxin.li@deakin.edu.au}
}

\maketitle

\blfootnote{*Hongzhi Yin is the corresponding author.}

\begin{abstract}
Latent factor models are the dominant backbones of contemporary recommender systems (RSs) given their performance advantages, where a unique vector embedding with a fixed dimensionality (\eg 128) is required to represent each entity (commonly a user/item). Due to the large number of users and items on e-commerce sites, the embedding table is arguably the least memory-efficient component of RSs. For any lightweight recommender that aims to efficiently scale with the growing size of users/items or to remain applicable in resource-constrained settings, existing solutions either reduce the number of embeddings needed via hashing, or sparsify the full embedding table to switch off selected embedding dimensions. However, as hash collision arises or embeddings become overly sparse, especially when adapting to a tighter memory budget, those lightweight recommenders inevitably have to compromise their accuracy. 
To this end, we propose a novel compact embedding framework for RSs, namely \textbf{C}ompositional \textbf{E}mbedding with \textbf{R}egularized \textbf{P}runing (CERP). Specifically, CERP represents each entity by combining a pair of embeddings from two independent, substantially smaller meta-embedding tables, which are then jointly pruned via a learnable element-wise threshold. In addition, we innovatively design a regularized pruning mechanism in CERP, such that the two sparsified meta-embedding tables are encouraged to encode information that is mutually complementary. Given the compatibility with agnostic latent factor models, we pair CERP with two popular recommendation models for extensive experiments, where results on two real-world datasets under different memory budgets demonstrate its superiority against state-of-the-art baselines. The codebase of CERP is available in \url{https://github.com/xurong-liang/CERP}.

\end{abstract}


\section{Introduction} \label{sec:intro}
The invention of recommender systems (RSs) greatly eases the difficulty of identifying and suggesting useful information or products from the sheer volume of data based on users' preferences. Most RSs leverage collaborative filtering through latent factor models, in which all entities (\ie users and items in most RSs) are  mapped to distinct, real-valued dense vectors of a unified dimension. Then, based on these vector representations, \ie embeddings, a pairwise similarity function (\eg dot product \cite{rendle_factorization_2010}, multi-layer perceptrons \cite{he_neural_2017}, graph neural networks \cite{he_lightgcn_2020}, \etc) can be learned to rank each item's relevance to a user. In latent factor-based collaborative filtering, all entities' embeddings are hosted in an embedding table and can be efficiently drawn via a look-up operation. 

Given the large number of possible users and items in recommendation services, the embedding table is commonly the heaviest component in an RS in terms of parameter sizes \cite{kang2020learning,nguyen2017argument,sun2020generic, zheng2023mmkgr,xia2022device,shi_compositional_2020}. Recently, with the frequently intersected needs for handling large-scale e-commerce data and deploying RSs on resource-constrained devices \cite{chen_learning_2021}, the memory consumption of embedding tables has become the major bottleneck that prevents RSs from scaling up. Take an example of the \emph{Amazon Product Reviews} dataset \cite{he2016ups} which includes $20.98$ million users and $9.35$ million items. If the embedding dimension is $128$, representing all these entities in a full embedding table incurs approximately $3.9$ billion parameters, translating into $31.2$ GB memory consumption for a double floating-point system. In comparison, the number of parameters used in the recommendation layer is almost negligible even for state-of-the-art RSs built upon deep neural networks (DNNs). Clearly, storing embedding vectors with a fixed dimension for all entities drastically escalates memory usage, making it intractable for RSs to scale to large datasets or support on-device applications. To this end, the urge for a lightweight recommender, created by utilizing a memory-efficient embedding structure, is raised. One \naive solution is to choose a small dimension size for all entities so that a low memory budget can be met. However, as the dimension size determines the ability to encode each entity's information \cite{zhao_autoemb_2020}, this approach heavily impedes the expressiveness of embedding vectors and thus, the recommendation accuracy.

To counter the inflexibility of fixed-size embeddings, one mainstream of recent memory-efficient RSs is to dynamically allocate different embedding sizes to entities. This is done by either constructing an automated search procedure to find the best embedding size for each entity from a set of predefined options \cite{ginart2021mixed, joglekar_neural_2020, zhao_autoemb_2020, liu_automated_2020}, or applying sparsification (\ie pruning) on the full embedding table to zero-out less important dimensions in every entity embedding \cite{liu_learnable_2021, chen_learning_2021, qu2022single,lyu_optembed_2022}. Though introducing varying dimensions helps selectively preserve embedding expressiveness for important entities (\eg popular items) when working toward a tight memory budget, the usable embedding dimensions in both search- and pruning-based approaches will decrease dramatically. Consequently, a substantial amount of embedded information is lost, sacrificing the accuracy when calculating user-item similarity. The key reason is that, these methods follow the conventional embedding table scheme, where every entity is still explicitly mapped to a unique embedding vector and no parameter sharing across different entities is allowed. Hence, existing dynamic embedding size allocation methods are essentially committed to reducing the average embedding dimension. However, the embedding parameter size is commonly dominated by the number of entities rather than the embedding dimension (\eg a million users and items \vs an embedding size of 128), the average dimension for each entity embedding has to drop significantly to meet a given memory budget.

\begin{figure*}[t!]
    \centering
    \begin{subfigure}[b]{\textwidth}
        \centering
        \includegraphics[width=.8\textwidth]{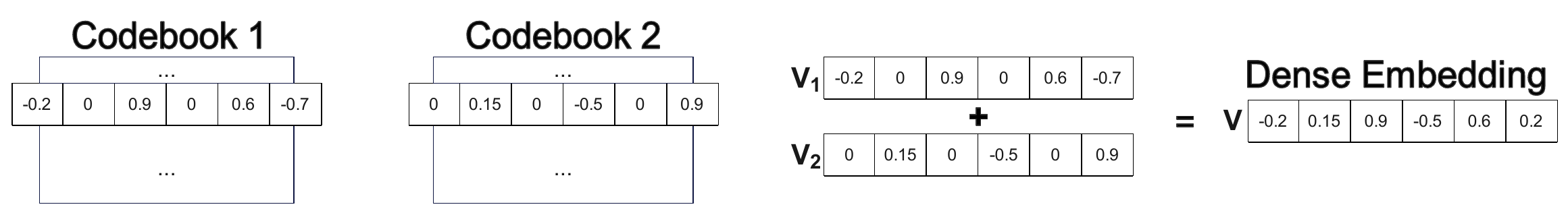}
        \caption[]%
            {{\small The deactivated dimensions in $\mathbf{v}_1$, $\mathbf{v}_2$ complement each other, leading to a dense compositional embedding with higher expressiveness.}}    
        \label{fig:dense_composition_emb}
    \end{subfigure}
    \begin{subfigure}[b]{\textwidth}
        \centering
        \includegraphics[width=.8\textwidth]{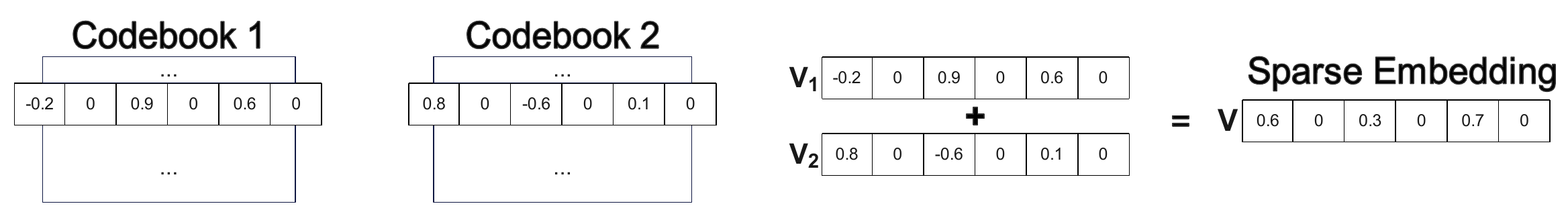}
        \caption[]%
            {{\small The deactivated dimensions in $\mathbf{v}_1$, $\mathbf{v}_2$ fully overlap, leading to a sparse compositional embedding with lower expressiveness.}}    
        \label{fig:sparse_composition_emb}
    \end{subfigure}
    \vspace*{-3mm}
    \caption{Illustration of the impact of complementary behavior between the $\mathbf{v}_1$ and $\mathbf{v}_2$.}
    \label{fig:complement_emb_intro}
    \vspace*{-5mm}
\end{figure*}

Naturally, another line of research in lightweight RSs is to enable parameter sharing to lower the number of embedding vectors needed for representing all entities, and hashing-based methods \cite{shi_compositional_2020, zhang_model_2020, li_lightweight_2021} are the most representative ones. In a nutshell, those solutions need one or several meta-embedding tables (\aka codebooks) consisting of fixed-size embedding vectors. By hashing each user/item ID into a combination of indexes, an entity embedding can be composed by merging all meta-embedding vectors (\eg via sum pooling) drawn from these hashed indexes, which is also termed a \textit{compositional embedding}. As such, the meta-embedding tables need to carry far fewer embedding vectors than a full embedding table, while still producing distinct representations for all entities. One example is that 10 meta-embeddings with dual hashing \cite{zhang_model_2020} can represent up to $C^{10}_2=45$ entities, with only $22\%$ of the parameters needed. While hashing-based methods can provide dense and fixed-size embeddings for all entities, the number of meta-embeddings allowed must be reduced for lower memory budgets. As a result, one meta-embedding has to be reused for a large number of compositional embeddings, \eg each meta-embedding will appear in over one-fifth of the compositional embeddings in the previous example, diluting the uniqueness of information they carry and eventually hurting the recommendation effectiveness \cite{kang_learning_2021}. Moreover, given a target memory budget, hash collisions are inevitable when the combinations of meta-embeddings are exhausted for the entity size. In this case, many entities are forced to share one identical compositional embedding \cite{kang_learning_2021}. Compared with dynamic embedding size allocation that is prone to producing excessively compact embeddings, hashing-based methods bear the risk of weakening the distinguishability of entity embeddings, which also impairs the utility of entity embeddings in ranking tasks. 

To this end, we put forward our lightweight embedding framework for RSs, namely \textbf{C}ompositional \textbf{E}mbedding with \textbf{R}egularized \textbf{P}runing (\textbf{CERP}). In CERP, instead of altering the usable embedding dimension or the number of total embeddings alone, we deploy a compositional embedding paradigm with two balanced codebooks, and design a pruning process to simultaneously sparsify them. On the one hand, with the same amount of parameters budgeted, sparsification allows to selectively switch off (\ie zero-out) less informative dimensions for each meta-embedding, which essentially squeezes out additional parameters to allow more meta-embeddings to be used. Thus, each meta-embedding is then less frequently reused in different compositional embeddings, uplifting the uniqueness of entity representations. On the other hand, as pruning now starts from the inherently smaller codebooks rather than the full embedding table, a given memory budget can be met with far fewer pruned parameters for every meta-embedding. In return, this further brings substantially denser and more expressive compositional embeddings for all entities. 
However, another challenge associated with the pruning process has to be tackled before CERP can enjoy benefits from both ends. 
Given the compositional nature of entity embeddings, the behavior of the pruning algorithm has to be regularized to avoid imbalanced or homogenized sparsification patterns between two codebooks. Taking a pair of meta-embeddings $\mathbf{v}_1, \mathbf{v}_2 \in \mathbb{R}^d$ as an example, with the commonly used sum pooling $\mathbf{v} = \mathbf{v}_1 + \mathbf{v}_2$, if half of their dimensions are deactivated, then the number of usable/non-zero embedding dimensions in the composed embedding $\mathbf{v}$ will range from $\frac{d}{2}$ to $d$. Ideally, if the deactivated dimensions in $\mathbf{v}_1$ and $\mathbf{v}_2$ complement each other, then the density of $\mathbf{v}$ is maximized (see Figure \ref{fig:dense_composition_emb}). On the contrary, as demonstrated in Figure \ref{fig:sparse_composition_emb}, if $\mathbf{v}_1$ fully overlaps with $\mathbf{v}_2$ on deactivated dimensions, despite consuming the same amount of $d$ parameters in $\mathbf{v}_1$ and $\mathbf{v}_2$, half of the dimensions in the resultant $\mathbf{v}$ do not carry any useful information, hence still being highly sparse. 
To alleviate this, we further design a pruning regularizer to facilitate complementary pruning so that the average size of entities' final embedding vectors is not compromised amid robust pruning. We summarize our main contributions below:
\begin{itemize}
    \item We innovatively propose to let the dynamic embedding size allocation and compositional embeddings complement each other, so as to facilitate memory-efficient and accurate lightweight recommendation.
    \item We design an embedding optimization framework CERP that utilizes two codebooks for generating compositional entity representations, where we further propose a novel pruning regularizer that coordinates joint pruning on the two codebooks.
    \item We conduct extensive experiments on real-world datasets to compare the performance of CERP with state-of-art embedding optimization methods. The results indicate that CERP achieves promising recommendation performance under strict memory budgets.
\end{itemize}

\section{Related Work} \label{sec:related_work}
Current literature attempts to construct lightweight recommender systems by considering various approaches, but with a unanimous focus on the compression of the embedding layer. In this section, we analyze different existing technical pathways toward this goal.

\textbf{Binary Code Representation Learning.} Since the conventional embedding table stores entity embeddings as real-valued dense vectors, which consumes significant storage space, early approaches target this by representing the embedding vectors using binary codes \cite{zhou2012learning, lian2017discrete, zhang2018discrete, zhang2020deep, lian_lightrec_2020}. There is also a branch that adaptively learns entities' binary hash code representations from real-valued dense embeddings \cite{kang2019candidate, tan2020learning} via approximation. Although binarized embedding vectors utilize much less space, the binarization process often causes quantization loss \cite{zhang_discrete_2016,liu2018discrete}, vastly distorting the entity representations and severely hurting the recommendation performance. 

\textbf{Automated Embedding Size Search.} A large amount of work \cite{ginart2021mixed, joglekar_neural_2020, zhao_autoemb_2020, liu_automated_2020,qu2023continuous,liu_learnable_2021, chen_learning_2021, lyu_optembed_2022} relies on automated machine learning (AutoML) for embedding optimization due to its convenience in training \cite{zheng_automl_2023}. MDE \cite{ginart2021mixed} takes in human-defined heuristics for entity embedding dimension assignment. Inspired by Neural Architecture Search (NAS) \cite{elsken2019neural}, techniques that conduct automatic embedding selection from pre-defined search space \cite{joglekar_neural_2020, cheng2020differentiable} were proposed. AutoEmb \cite{zhao_autoemb_2020} and ESAPN \cite{liu_automated_2020} were further invented to learn suitable embedding dimensions based on the popularity of entities. AutoEmb \cite{zhao_autoemb_2020} devises soft selection to express embedding vectors as the weighted sum of multiple embedding sizes. ESAPN \cite{liu_automated_2020} designs an automatic reinforcement learning (RL) agent for embedding size selection. Although these dimension search methods may find the appropriate dimension size for each entity, the slow training speed of NAS \cite{pham2018efficient} throttles their training efficiency. In addition, for a method that uses reinforcement learning for dimension search, the search space will normally be enormous. Thus, increases the difficulty for the learning agent to find the most optimal embedding sizes. 

\textbf{Embedding Pruning.} As another alternative, the main idea of pruning-based methods is to automatically learn the importance of parameters in embeddings or components in models \cite{zheng_automl_2023}. The unimportant or redundant ones are then pruned (\ie deleted) to lower memory consumption. Techniques that belong to this category are PEP \cite{liu_learnable_2021} and OptEmbed \cite{lyu_optembed_2022}. PEP \cite{liu_learnable_2021} applies $L_1$ regularization on the full embedding table directly to increase the sparsity adaptively via a learnable pruning threshold. While the robustness of $L_1$ regularization in sparsification \cite{ramirez2013l1}, amid the tight memory budget and the fact that each entity is assigned a unique embedding, the number of usable dimension drops drastically, causing a loss of embedding fidelity and sacrificing system accuracy. OptEmbed \cite{lyu_optembed_2022} combines a trainable pruning threshold with a one-shot NAS dimension search to perform both row-wise and column-wise embedding optimization. However, the time-consuming nature of NAS \cite{pham2018efficient} remains a bottleneck for its scalability.

\textbf{Compositional Embeddings.} Compared with pruning, compositional embeddings \cite{weinberger_feature_2009, zhang_model_2020, shi_compositional_2020, li_lightweight_2021, kang_learning_2021, desai2021semantically} are an ideal solution to lower the number of embedding rows, and meanwhile preserving dense embeddings. The main idea is to represent entities with a combination of meta-embedding vectors, where the meta-embeddings for entities are determined by hash functions and can be shared \cite{weinberger_feature_2009}. However, a high memory constraint will also limit the number of usable meta-embeddings, bringing the risk of hash collisions. LCE \cite{hang2022lightweight} only explicitly trains user (or item) embeddings and applies composition operator and GNN to infer item (or user) embeddings in real time. However, this technique does not avoid memory exhaustion when handling an extreme number of users/items. Kang \etal \cite{kang_learning_2021} replace the embedding tables with a DNN-based embedding generator fed with carefully crafted hash codes for each user/item. Nevertheless, the quality of the generated embeddings relies on excessively long hash codes (\eg 1,024 as in \cite{kang_learning_2021}) to compensate for the limited expressiveness of the DNN. It is also worth noting that there is another direction \cite{wang_next_2020,xia2023efficient, yin_nimble_2022, xia2022device} to utilize tensor train (TT) decomposition \cite{oseledets2011tensor} to compress the embedding table as a sequence of TT-cores, which can be interpreted as a special case of compositional embeddings \cite{chen_learning_2021}. However, given the widely acknowledged computational overheads introduced by the sequential matrix operations \cite{yin_nimble_2022}, it is a less practical solution for large-scale RSs.

\section{Compositional Embedding with\\ Regularized Pruning} \label{sec:method}
We present CERP, our proposed lightweight embedding approach for recommendation in this section by introducing the detailed design of its components. 

\subsection{Generating Compositional Embeddings} \label{Sec:com_embedding_structure}
In a typical latent factor recommender, an entity (\ie user/item) is represented by a distinct, $d$-dimensional embedding vector, contributing to an embedding table with $(|\mathcal{U}|+|\mathcal{I}|) \times d$ parameter consumption. $\mathcal{U}$ and $\mathcal{I}$ are the sets of all users and all items, respectively. As described in Section \ref{sec:intro}, to reduce the size of the full embedding table, each entity embedding in CERP is composed by combining a pair of meta-embeddings $\mathbf{p}$, $\mathbf{q}\in \mathbb{R}^{d}$ respectively drawn from two smaller embedding tables $\mathbf{P}$, $\mathbf{Q}\in \mathbb{R}^{b \times d}$ (i.e., codebooks). Each row in $\mathbf{P}$ and  $\mathbf{Q}$ corresponds to a meta-embedding, and $b$ is the number of meta-embeddings in each codebook, also termed \textit{bucket size}. For each entity, to generate its compositional embedding, we first retrieve one meta-embedding vector from each codebook, and then merge both vectors. In this case, a total of $b^2$ different combinations of $(\mathbf{p}, \mathbf{q})$ can be guaranteed. As our setting allows $2b \ll |\mathcal{U}|+|\mathcal{I}|$ by several magnitudes while ensuring $b^2 \geq |\mathcal{U}|+|\mathcal{I}|$, such a compositional paradigm can significantly cut the parameter consumption while preserving the uniqueness of all entity embeddings. 

If a full embedding table is in use, then each entity's associated index $k \in [0, N-1]$ will point to its embedding stored at the corresponding row of the full embedding table. In our compositional embedding scheme, each entity now needs to have a pair of indexes $k_p,k_q \in [0, b-1]$ for the two codebooks. Intuitively, this can be accomplished by applying some hash functions to map the original entity index $k$ to $(k_p,k_q)$. Meanwhile, in a recommendation setting, to obtain optimal expressiveness of the resulted entity embeddings, each meta-embedding needs to be prevented from being frequently reused for composing entity embeddings. Hence, we would also like to spread the hashed values as evenly as possible. To achieve this, we design a balanced hashing trick to assign each entity a unique combination without introducing additional learnable model parameters. Let all users/items be indexed with continuous integers $k\in[0,|\mathcal{U}|+|\mathcal{I}|-1]$, then the two hashed indexes are computed via:
\begin{equation}
    \begin{aligned}
        k_p &=k \!\!\!\!\mod b,\\
        k_q &=k \,\,\mathrm{ div }  \ceil*{\frac{|\mathcal{U}| + |\mathcal{I}|}{b}},
    \end{aligned}
    \label{eq:hashing}
\end{equation}
where $\mathrm{mod}$ and $\mathrm{div}$ are respectively the modulo and integer division operators. 
Essentially, with Equation \ref{eq:hashing}, each $k_p$/$k_q$ value only appears $\ceil*{\frac{|\mathcal{U}|+|\mathcal{I}|}{b}}$ times in all compositional embeddings. For the $k$-th entity, we identify the $k_p$-th and $k_q$-th meta-embeddings respectively from $\mathbf{P}$ and $\mathbf{Q}$, and compute its compositional embedding via sum pooling:
\begin{equation}
\begin{aligned}
    \mathbf{e}_k &= \mathbf{p} + \mathbf{q},\\
    \mathbf{p}&=\mathbf{P}[k_p]^{\top}, \,\,\, \mathbf{q}=\mathbf{Q}[k_q]^{\top}.
\end{aligned}
\end{equation}
Considering that $\mathbf{p}$ and $\mathbf{q}$ will be heavily sparsified in the pruning stage, the use of sum pooling produces denser compositional embeddings, especially compared with other vector combinatory operations like element-wise product and concatenation that will lead to more zero-valued entries in the resulted embeddings. 

\subsection{Regularized Embedding Pruning} \label{Sec:pruning}
If the memory budget is sufficient and a decent bucket size $b$ for both codebooks is used, the uniqueness of information encoded in each compositional embeddings can be strengthened, since each meta-embedding will be shared by fewer entities. However, when the memory budget shrinks and a small $b$ has to be used, it inevitably lowers the expressiveness of all entity embeddings. Hence, in CERP, we propose to sparsify the codebooks via pruning, such that the codebooks can retain a relatively larger bucket size under each given memory budget. During pruning, the less informative dimensions in each meta-embedding are masked by zeros, where the resultant codebooks can be efficiently handled by the latest sparse matrix storage techniques \cite{sedaghati2015automatic,virtanen2020scipy} that bring negligible cost for storing zero-valued entries.

Compared with methods that reduce embedding dimensions by recursively searching for the best embedding size for each entity, pruning-based solutions have several advantages. Firstly, most search-based methods have to choose the optimal embedding size from a set of predefined discrete options (\eg $\{16,32,64,128\}$), making the final embedding sizes far less refined than pruning methods that can individually decide to block or keep each dimension of $\mathbb{R}^d$. Secondly, unlike pruning methods, search-based methods are heavily entangled with reinforcement learning due to the need for iterative search and evaluate different actions, which does not favor large-scale applications. Thirdly, while search-based methods will vary the dimensionality across different embeddings, all pruned meta-embeddings are still $d$-dimensional vectors that are partially masked, thus fully supporting the need for vector combinatory operations in compositional paradigms. 


\begin{figure*}[t!]
  \centering
  \includegraphics[width=.8\textwidth]{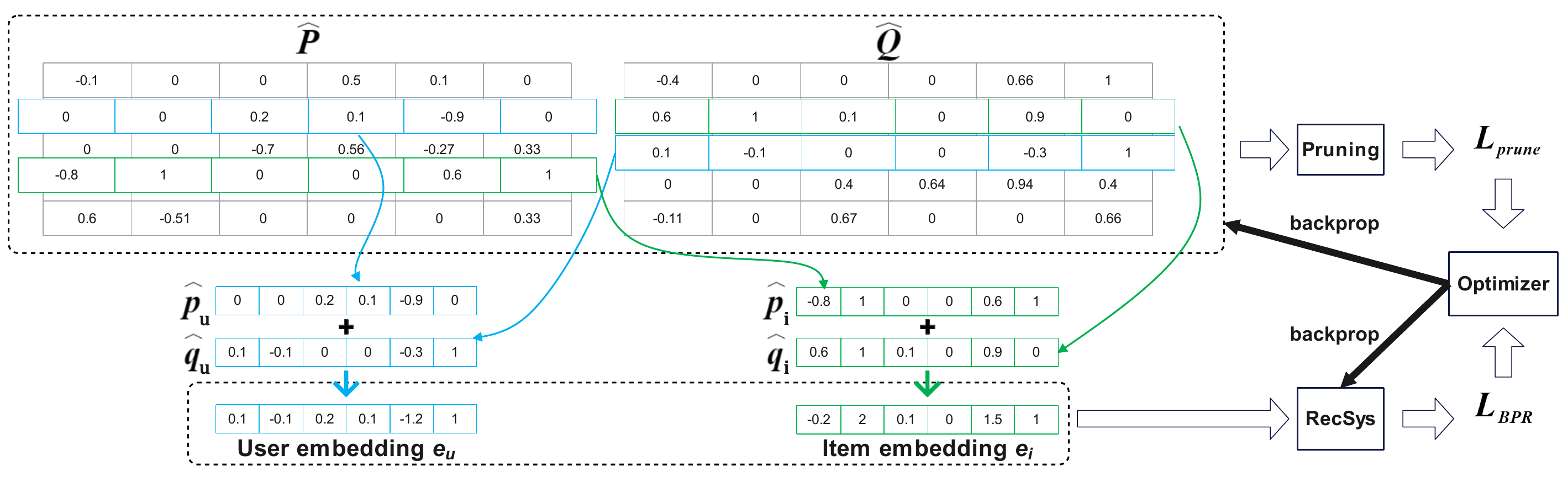}
  \vspace{-0.3cm}
  \caption{The overview of the main components and optimization process of CERP.}
  \label{Fig:overview}
  \vspace{-0.6cm}
\end{figure*}

Considering the codebooks $\mathbf{P}$, $\mathbf{Q}$, the pruning objective is formulated in conjunction with the overall loss $\mathcal{L}$ (\ie the recommendation loss with other optional side-task losses) as
\begin{equation}
    \min \; \mathcal{L}, \;\;\; s.t. \; ||\mathbf{P}||_0+||\mathbf{Q}||_0 \leq t, 
\label{eq:prune}
\end{equation} 
where $||\cdot||_0$ is the $L_0$-norm that counts the number of non-zero entries in a matrix, and $t \in \mathbb{N}$ is a predefined threshold indicating the maximum parameter number allowed. Unfortunately, straightforwardly optimizing Equation \ref{eq:prune} is intractable, given $L_0$-norm's non-convexity and the NP-hard nature of such a combinatorial problem \cite{kusupati_soft_2020}. A proper way to wrap around this is to perform $L_1$ regularization, in which the original $L_0$-norm problem is projected on $L_1$-ball to make the original problem end-to-end differentiable. Enlightened by the effectiveness of $L_1$ convex relaxation \cite{ramirez2013l1}, we reparameterize the pruning process \cite{kusupati_soft_2020, liu_learnable_2021} of two codebooks into the following functions to approximate the $L_0$-based sparsity constraint:
\begin{equation} \label{eq:pruning_formulas}
    \begin{aligned}
        \mathbf{\widehat{P}} &= \mathrm{prune}(\mathbf{P}, \mathbf{S}_p) = \mathrm{sign}(\mathbf{P}) \odot \mathrm{ReLU}(|\mathbf{P}| - \sigma(\mathbf{S}_p)),\\
        \mathbf{\widehat{Q}} &= \mathrm{prune}(\mathbf{Q}, \mathbf{S}_q) = \mathrm{sign}(\mathbf{Q}) \odot \mathrm{ReLU}(|\mathbf{Q}| - \sigma(\mathbf{S}_q)),
    \end{aligned}
\end{equation}
where $\odot$ denotes element-wise multiplication, $\mathrm{ReLU}(\cdot)$ and $\sigma(\cdot)$ are respectively the rectified linear unit and sigmoid functions, and $\mathrm{sign}(\cdot)$ is the signum function that returns 1, -1, and 0 respectively for inputs above, below, or equal to 0. We introduce two learnable soft threshold matrices $\mathbf{S}_p, \mathbf{S}_q \in \mathbb{R}^{b\times d}$ to control the sparsity of both codebooks in a fine-grained, element-wise manner. Taking codebook $\mathbf{P}$ as an example, $\mathrm{ReLU}(|\mathbf{P}| - \sigma(\mathbf{S}_p))$ will zero-out entries in $\mathbf{P}$ if the corresponding threshold elements in $\mathbf{S}_p$ receive a large value, while the multiplication with $\mathrm{sign}(\mathbf{P})$ helps recover the positivity/negativity of retained elements to ensure expressiveness. Notably, once the soft thresholds are learned, only the computed sparse codebooks $\mathbf{\widehat{P}}$ and $\mathbf{\widehat{Q}}$ will be kept and deployed for the recommender, thus avoiding unnecessary parameter consumption. 

As training proceeds, the soft thresholds $\mathbf{S}_p$ and $\mathbf{S}_q$ are to be updated in every optimization iteration, so will both sparsified codebooks $\mathbf{\widehat{P}}$ and $\mathbf{\widehat{Q}}$ until the desired parameter number $t$ is met. However, it is worth mentioning that function $\mathrm{sign}(\cdot)$ is non-differentiable at zero and has a zero-valued gradient at all other points, so we take advantage of subgradients \cite{kusupati_soft_2020} to facilitate end-to-end backpropagation. 
Taking codebook $\mathbf{P}$ as an example, the subgradients for optimizing $\mathbf{P}$ and the soft threshold $\mathbf{S}_p$ are respectively the following:
\begin{equation}
\begin{aligned}
    \nabla _{\mathbf{P}} &= \frac{\partial \mathcal{L}(\mathcal{D}, \mathrm{prune}(\mathbf{P}, \mathbf{S}_p))}{\partial \mathrm{prune}(\mathbf{P}, \mathbf{S}_p)} \odot \mathbf{1}\{\mathrm{prune}(\mathbf{P}, \mathbf{S}_p)\neq 0\},\\
    \nabla _{\mathbf{S}_p} &= - \frac{\partial \sigma(\mathbf{S}_p)}{\partial \mathbf{S}_p} \cdot \bigg{\langle}\frac{\partial \mathcal{L}(\mathcal{D}, \mathrm{prune}(\mathbf{P}, \mathbf{S}_p))}{\partial \mathrm{prune}(\mathbf{P}, \mathbf{S}_p)} , \\ & \;\;\;\;\;\;\;\;\;\;\;\;\;\;\;\;\;\;\;\;\;\;\;\;\; \mathrm{sign}(\mathbf{P}) \odot \mathbf{1}\{\mathrm{prune}(\mathbf{P}, \mathbf{S}_p)\neq 0\}\bigg{\rangle},
    \end{aligned}
\end{equation}
where $\mathcal{D}$ is the full training set, $\langle\cdot, \cdot \rangle$ is the matrix inner product, and $\mathbf{1}\{\cdot\}$ is the indicator function that only keeps gradients for non-zero entries for gradient descent. Analogously, we can also derive the gradients for the other codebook $\mathbf{Q}$ and soft threshold $\mathbf{S}_q$.

Now, following the same hashing rule in Equation \ref{eq:hashing}, the entity embeddings are composed from the sparsified codebooks $\widehat{\mathbf{P}}$ and $\widehat{\mathbf{Q}}$, denoted by  $\mathbf{e}_k=\widehat{\mathbf{p}}+\widehat{\mathbf{q}}$. As stated in Section \ref{sec:intro}, given the compositional nature of entity embeddings, it is beneficial to enforce that the pruned dimensions in $\widehat{\mathbf{p}}$ and $\widehat{\mathbf{q}}$ are complementary rather than repetitive. Ideally, for all $k\in [0,|\mathcal{U}|+|\mathcal{I}|-1]$, we want to achieve this for every possible $(\mathbf{\hat{p}},\mathbf{\hat{q}})$ pair, leading to the following optimization goal:
\begin{equation}
\underset{\widehat{\mathbf{P}},\,\widehat{\mathbf{Q}}}{\mathrm{argmin}} - \sum_{k_p=1}^{b}\sum_{k_q=1}^{b}||\widehat{\mathbf{P}}[k_p]+\widehat{\mathbf{Q}}[k_q]||_0,
\label{eq:prune_reg_raw}
\end{equation}
which is essentially maximizing the total number of non-zero entries in all compositional entity embeddings. However, the non-convexity of $L_0$ norm and time-consuming iterative computation over $b^2$ combinations suggest that we need a more computationally feasible solution to regularize the pruning process. Thus, we design the following instance-level regularization term to approximate the desired pruning behavior in Equation \ref{eq:prune_reg_raw}:
\begin{equation} \label{eq:pruning_loss}
\mathcal{L}_{prune} = - \sum_{k\in \mathcal{B}} \| \tanh\left(\eta \mathbf{e}_k \right) \|^2_2
\end{equation}
where $\mathcal{B}$ is the current training batch. With a large positive temperature value $\eta$ (\eg $10^3$), for any non-zero entry $e\in \mathbf{e}_k$, we have $\tanh\left(\eta e\ \right)\approx -1$/$1$. 

\textbf{Rationale of The Proposed Pruning Regularizer.} 
To better motivate the design of $\mathcal{L}_{prune}$, we use a toy example 
to showcase the relationship between $\mathcal{L}_{prune}$ and different pruning behaviors on $\widehat{\mathbf{p}}$ and $\widehat{\mathbf{q}}$ that compose $\mathbf{e}_k$. 
In a nutshell, with the rescaled $\tanh(\cdot)$, the squared $L_2$ norm counts the number of non-zero entries in each $\mathbf{e}_k$. Despite the same amount of 6 parameters consumed 
they may correspond to three possible cases where the non-zero dimensions in two meta-embeddings completely overlap, partially complementary, and completely complementary. As the density of the resultant entity embedding grows, the squared norm returns a higher value, which translates into a lower $\mathcal{L}_{prune}$. Without the pruning regularizer, the sparsification focuses on lowering the parameter number only, and may result in suboptimal utility of the compositional entity embeddings. 
In contrast, $\mathcal{L}_{prune}$ is able to advocate minimizing the amount of zero-valued entries in $\mathbf{e}_k$, which in fact rewards a mismatching pattern in the pruned dimensions between every pair of $(\widehat{\mathbf{p}}, \widehat{\mathbf{q}})$. Furthermore, $\mathcal{L}_{prune}$ is fully differentiable\footnote{Because we design $\mathcal{L}_{prune}$ only for regularization purpose, instead of using a non-differentiable, hard selection function like $\mathrm{sign}(\cdot)$, our rescaled $\tanh(\cdot)$ is simple yet effective for counting non-zero entries.}, and can be efficiently appended to the mini-batch training process.

\subsection{Joint Optimization of Recommendation and Pruning}\label{sec:joint_train}
We provide an overview of the optimization process of CERP in Figure \ref{Fig:overview}. After obtaining sparsified codebooks $\widehat{\mathbf{P}}$ and $\widehat{\mathbf{Q}}$, we may retrieve the sparsified entity embeddings. For a (user, item) pair indexed by $(u, i)$, denoting the retrieved user and item embeddings as $\mathbf{e}_u, \mathbf{e}_i$, then the user-item affinity score for personalized item ranking is computed as follows:
\begin{equation}
    \hat{y}_{ui} = f_{rec}(\mathbf{e}_u, \mathbf{e}_i),
\end{equation}
where $f_{rec}(\cdot,\cdot)$ is the base recommender that takes the sparsified user and item embeddings as its input, and estimates the pairwise user-item similarity $\hat{y}_{ui}$. The choice of $f_{rec}(\cdot,\cdot)$ can be an arbitrary latent factor model that requires embeddings. Since we are optimizing the quality of personalized recommendation ranking, we adopt the Bayesian personalized ranking (BPR) loss \cite{rendle_bpr_2009} for model parameter learning:
\begin{equation} \label{eq:bprloss}
    \mathcal{L}_{BPR} = \sum_{(u, i^+, i^-) \in \mathcal{B}} -\ln{\sigma(\hat{y}_{ui^+} - \hat{y}_{ui^-})},
\end{equation}
where $\mathcal{B}$ is a training batch of the full dataset, and triplets $(u, i^+, i^-)$ contains the sampled user $u$'s positively rated item $i^+$ and negatively rated/unvisited item $i^-$.
For simplicity, we omit the $L_2$ penalization which is a common practice for preventing overfitting.

Finally, we define the joint loss equation to facilitate simultaneous optimization of both the recommendation and sparsification tasks:
\begin{equation} \label{eq:total_loss_prune}
    \mathcal{L} = \mathcal{L}_{BPR} + \gamma \mathcal{L}_{prune},
\end{equation}
where $\gamma \in [0,1]$ adjusts the weight of $\mathcal{L}_{prune}$ in the total loss $\mathcal{L}$. The higher the value of $\gamma$, the stronger the constraint will be for the pruning behavior of CERP, hence a slower pruning process may be observed. To coordinate pruning efficiency and embedding quality simultaneously, we additionally impose an exponential decay on $\gamma$. 

\textbf{Rationale of Exponential Decay on $\gamma$.}
We first set $\gamma$ to an initial value, then by the end of each pruning epoch, $\gamma$ is reduced by half. The motivation for having a gradually decreasing $\gamma$ value is that, at an early stage of the pruning process, we aim to retain valuable embedding elements so the embedding quality is preserved. When the embedding table becomes highly sparse as the pruning goes, instead of overly emphasizing where to prune, it is more beneficial to focus on optimizing the information stored in each sparse embedding by lowering the impact of $\mathcal{L}_{prune}$.

From Figure \ref{Fig:overview}, we may see that the recommendation loss and pruning regularization loss are optimized at the same time. The joint training is terminated when the total parameter consumption of the two sparsified codebooks reaches the target sparsity rate $s$:
\begin{displaymath}
    s = \frac{||\widehat{\mathbf{P}}||_0 + ||\widehat{\mathbf{Q}}||_0}{(|\mathcal{U}|+|\mathcal{I}|)d} \times 100\%,
\end{displaymath}
which is the ratio between the final embedding parameter consumption and the size of a full embedding table with $|\mathcal{U}|+|\mathcal{I}|$ $d$-dimensional vectors.

\subsection{Parameter Retraining}
Work that utilizes pruning as the embedding optimization strategy typically requires a parameter retraining stage after the pruning stage is conducted \cite{liu_learnable_2021, lyu_optembed_2022} in order to prevent the interference on embedding value learning caused by pruning itself. We follow this fashion in our work as well. When the joint optimization described above is done, we can obtain the fixed pruning masks from both sparsified codebooks:
\begin{equation}
        \mathbf{\widehat{P}}_{mask} = \lvert \mathrm{sign}(\mathbf{\widehat{{P}}}) \rvert, \,\,\,\, \mathbf{\widehat{Q}}_{mask} = \lvert \mathrm{sign}(\mathbf{\widehat{{Q}}}) \rvert.
\end{equation}
It is clear that $\mathbf{\widehat{P}}_{mask}, \mathbf{\widehat{Q}}_{mask} \in \{0, 1\}^{b \times d}$ record whether an element in $\mathbf{{P}}$ and $\mathbf{{Q}}$ should be pruned or not. 
Then, element-wise multiplication is performed with $\mathbf{\widehat{P}}_{mask}$ and $\mathbf{\widehat{Q}}_{mask}$ respectively to mask out deactivated codebook entries during retraining. The retraining is summarized as the following:
\begin{equation}
    \begin{aligned}
        \underset{\mathbf{{P}}, \mathbf{Q}, \Theta}{\mathrm{argmin}} \,\mathcal{L}_{BPR}(\mathcal{D}, \mathbf{{P}}  \odot \mathbf{\widehat{P}}_{mask}, \mathbf{{Q}}  \odot \mathbf{\widehat{Q}}_{mask}),
    \end{aligned}
\end{equation}
where $\Theta$ denotes all trainable parameters in the recommender $f(\cdot,\cdot)$. When the retraining converges, only the sparse codebooks $\widehat{\mathbf{{P}}} = \mathbf{{Q}}  \odot \mathbf{\widehat{Q}}_{mask}$ and $\widehat{\mathbf{{Q}}} = \mathbf{{Q}}  \odot \mathbf{\widehat{Q}}_{mask}$ need to be kept as the embedding component for the recommendation model.

\section{Experiments} \label{sec:experiments}
In this section, we organize experiments to evaluate the effectiveness of CERP. Specifically, we are interested in answering the following research questions (RQs):
\begin{itemize}
    \item \textbf{RQ1:} Does our framework work well compared to other baselines under various memory budgets?
    \item \textbf{RQ2:} What is the effect of CERP's key components?
    \item \textbf{RQ3:} How do different hyperparameter settings affect CERP's performance?
\end{itemize}

\subsection{Experimental Settings}

\subsubsection{Datasets}
We use two publicly available benchmark datasets: \textbf{Gowalla} and \textbf{Yelp2020}. Gowalla dataset is available on LightGCN's \cite{he_lightgcn_2020} official code repository\footnote{\url{https://github.com/gusye1234/LightGCN-PyTorch}}, while Yelp2020 can be found on HGCF's \cite{sun2021hgcf} official code repository\footnote{\url{https://github.com/layer6ai-labs/HGCF}}. The detailed statistics of both datasets are summarized in Table~\ref{tab:dataset_stat}. We adopt train/test/validation splits protocol in \cite{wang_neural_2019} to randomly select 80\% of interactions as the train set. The test set contains the remaining 20\% of interactions and 10\% of the train set was further picked to form the validation set. For each user-positive item interaction, we sample 5 negative items.

\begin{table}[h]
    \caption{Statistics of datasets used in our work.}
    \label{tab:dataset_stat}
    \centering
        \vspace*{-1mm}

    \begin{tabular}{c  c c  c c}
        \hline
        Dataset & \#User & \#Item & \#Interaction & Density\\
        \hline
        Gowalla & 29,858 & 40,981 & 1,027,370 & 0.084\%\\
        Yelp2020  & 71,135  & 45,063 & 1,782,999 & 0.056\%\\
        \hline
    \end{tabular}
\end{table}

\begin{table*}[t!]
    \caption{Performance comparison $w.r.t.$ different embedding sparsity rates, where ``Avg Dim''  denotes the average of the actual embedding dimension achieved by each method. Percentages indicate the performance difference between CERP and the best baseline results. We use bold font to highlight the best result under each target sparsity rate $s$.}
    \label{tab:overall_performance}
    \vspace*{-2mm}
\centering
\resizebox{1.01\width}{!}{
\setlength\tabcolsep{1.2pt}
\begin{tabular}{c|cccccccc||cccccccc}
\hline
         & \multicolumn{8}{c||}{Gowalla}                                                                                                                                                                                                                               & \multicolumn{8}{c}{Yelp2020}                                                                                                                                                                                                                        \\ \cline{2-17}
         & \multicolumn{4}{c|}{MLP}                                                                                                               & \multicolumn{4}{c||}{LightGCN}                                                                                     & \multicolumn{4}{c|}{MLP}                                                                                                        & \multicolumn{4}{c}{LightGCN}                                                                                      \\ \cline{2-17}
Method   & Sparsity              & \begin{tabular}[c]{@{}c@{}}Avg\\Dim\end{tabular} & N@10            & \multicolumn{1}{c|}{R@10}            & Sparsity              & \begin{tabular}[c]{@{}c@{}}Avg\\Dim\end{tabular} & N@10            & R@10            & Sparsity              & \begin{tabular}[c]{@{}c@{}}Avg\\Dim\end{tabular} & N@10            & \multicolumn{1}{c|}{R@10}     & Sparsity              & \begin{tabular}[c]{@{}c@{}}Avg\\Dim\end{tabular} & N@10            & R@10            \\ \hline
ESAPN    & 87.89\%               & 15.5                                                  & 0.0045          & \multicolumn{1}{c|}{0.0051}          & 89.14\%               & 13.9                                                  & 0.0282          & 0.0292          & 89.83\%               & 13.02                                                 & 0.0035          & \multicolumn{1}{c|}{0.0040}   & 90.35\%               & 12.35                                                 & 0.0064          & 0.0078          \\ \hline
AutoEmb  & 89.84\%               & 13                                                    & 0.0013          & \multicolumn{1}{c|}{0.0012}          & 89.84\%               & 13                                                    & 0.0275          & 0.0275          & 89.84\%               & 13                                                    & 0.0003          & \multicolumn{1}{c|}{0.0003}   & 89.84\%               & 13                                                    & 0.0073          & 0.0090          \\ \hline
OptEmbed & 89.97\%               & 12.84                                                 & 0.0276          & \multicolumn{1}{c|}{0.0276}          & 89.34\%               & 13.64                                                 & 0.0373          & 0.0369          & 88.31\%               & 14.96                                                 & 0.0055          & \multicolumn{1}{c|}{0.0068}   & 87.96\%               & 15.41                                                 & 0.0084          & 0.0089          \\ \hline
DHE      & 90.63\%               & 12                                                    & 0.0081          & \multicolumn{1}{c|}{0.0069}          & 90.63\%               & 12                                                    & 0.0046          & 0.0050          & 90.63\%               & 12                                                    & 0.0017          & \multicolumn{1}{c|}{0.0023}   & 90.63\%               & 12                                                    & 0.0013          & 0.0016          \\ \hline
UD       & \multirow{5}{*}{90\%} & 12                                                    & 0.0265          & \multicolumn{1}{c|}{0.0266}          & \multirow{5}{*}{90\%} & 12                                                    & 0.0875          & 0.0797          & \multirow{5}{*}{90\%} & 12                                                    & 0.0049          & \multicolumn{1}{c|}{0.0052}   & \multirow{5}{*}{90\%} & 12                                                    & 0.0211          & 0.0251          \\
PEP      &                       & 12.79                                                 & 0.0251          & \multicolumn{1}{c|}{0.0252}          &                       & 12.79                                                 & 0.0685          & 0.0607          &                       & 12.80                                                 & 0.0036          & \multicolumn{1}{c|}{0.0046}   &                       & 12.78                                                 & 0.0157          & 0.0186          \\
QR       &                       & 128                                                   & 0.0288          & \multicolumn{1}{c|}{0.0273}          &                       & 128                                                   & 0.0776          & 0.0724          &                       & 128                                                   & 0.0054          & \multicolumn{1}{c|}{0.0062}   &                       & 128                                                   & 0.0193          & 0.0225          \\
CERP     &                       & 124.33                                                & \textbf{0.0312} & \multicolumn{1}{c|}{\textbf{0.0288}} &                       & 125.39                                                & \textbf{0.0965} & \textbf{0.0926} &                       & 122.94                                                & \textbf{0.0061} & \multicolumn{1}{c|}{\textbf{0.0067}}   &                       & 125.79                                                & \textbf{0.0230} & \textbf{0.0267} \\
         &                       &                                                       & (+8.3\%)        & \multicolumn{1}{c|}{(+5.7\%)}        &                       &                                                       & (+10.3\%)       & (+16.2\%)       &                       &                                                       & (+12.7\%)       & \multicolumn{1}{c|}{(+8.0\%)} &                       &                                                       & (+9.0\%)        & (+6.4\%)        \\ \hline
UD       & \multirow{5}{*}{95\%} & 6                                                     & 0.0248          & \multicolumn{1}{c|}{0.0257}          & \multirow{5}{*}{95\%} & 6                                                     & 0.0617          & 0.0568          & \multirow{5}{*}{95\%} & 6                                                     & 0.0042          & \multicolumn{1}{c|}{0.0048}   & \multirow{5}{*}{95\%} & 6                                                     & 0.0161          & 0.0190          \\
PEP      &                       & 6.39                                                  & 0.0277          & \multicolumn{1}{c|}{0.0268}          &                       & 6.39                                                  & 0.0664          & 0.0589          &                       & 6.40                                                  & 0.0052          & \multicolumn{1}{c|}{0.0055}   &                       & 6.39                                                  & 0.0111          & 0.0121          \\
QR       &                       & 128                                                   & 0.0280          & \multicolumn{1}{c|}{0.0254}          &                       & 128                                                   & 0.0598          & 0.0546          &                       & 128                                                   & 0.0056          & \multicolumn{1}{c|}{0.0063}   &                       & 128                                                   & 0.0151          & 0.0166          \\
CERP     &                       & 83.96                                                 & \textbf{0.0307} & \multicolumn{1}{c|}{\textbf{0.0288}} &                       & 84.88                                                 & \textbf{0.0913} & \textbf{0.0877} &                       & 80.85                                                 & \textbf{0.0057} & \multicolumn{1}{c|}{\textbf{0.0067}}   &                       & 82.37                                                 & \textbf{0.0206} & \textbf{0.0236} \\
         &                       &                                                       & (+9.6\%)        & \multicolumn{1}{c|}{(+7.1\%)}        &                       &                                                       & (+37.6\%)       & (+48.8\%)       &                       &                                                       & (+1.7\%)        & \multicolumn{1}{c|}{(+6.0\%)} &                       &                                                       & (+28.0\%)       & (+24.3\%)       \\ \hline
UD       & \multirow{5}{*}{99\%} & 1                                                     & 0.0273          & \multicolumn{1}{c|}{0.0261}          & \multirow{5}{*}{99\%} & 1                                                     & 0.0416          & 0.0406          & \multirow{5}{*}{99\%} & 1                                                     & 0.0043          & \multicolumn{1}{c|}{0.0050}   & \multirow{5}{*}{99\%} & 1                                                     & 0.0092          & 0.0104          \\
PEP      &                       & 1.28                                                  & 0.0264          & \multicolumn{1}{c|}{0.0277}          &                       & 1.28                                                  & 0.0423          & 0.0346          &                       & 1.28                                                  & 0.0052          & \multicolumn{1}{c|}{0.0054}   &                       & 1.28                                                  & 0.0042          & 0.0052          \\
QR       &                       & 128                                                   & 0.0222          & \multicolumn{1}{c|}{0.0212}          &                       & 128                                                   & 0.0575          & 0.0540          &                       & 128                                                   & 0.0058          & \multicolumn{1}{c|}{0.0067}   &                       & 128                                                   & 0.0071          & 0.0081          \\
CERP     &                       & 18.30                                                 & \textbf{0.0298} & \multicolumn{1}{c|}{\textbf{0.0279}} &                       & 17.99                                                 & \textbf{0.0831} & \textbf{0.0816} &                       & 18.66                                                 & \textbf{0.0061} & \multicolumn{1}{c|}{\textbf{0.0071}}   &                       & 18.61                                                 & \textbf{0.0167} & \textbf{0.0190} \\
         &                       &                                                       & (+9.0\%)        & \multicolumn{1}{c|}{(+0.7\%)}        &                       &                                                       & (+44.5\%)       & (+51.1\%)       &                       &                                                       & (+5.1\%)        & \multicolumn{1}{c|}{(+5.6\%)} &                       &                                                       & (+82.3\%)       & (+83.1\%)       \\ \hline
\end{tabular}
}
\vspace{-0.4cm}
\end{table*}

\subsubsection{Backbone Recommenders and Baselines}
CERP is model-agnostic to various latent factor recommenders for lowering the memory consumption of their embedding tables. To comprehensively evaluate our method's efficacy and generalizability across different base recommenders, we pair CERP with two recommenders, namely the multi-layer percerptron (MLP) from neural collaborative filtering \cite{he_neural_2017} (\ie its deep component), and the light graph convolution network (LightGCN) \cite{he_lightgcn_2020}, which are commonly used backbones \cite{chen2020try,chen2022thinking,yu2023self}. Specifically, both recommenders will have their full embedding table replaced by the sparsified one in CERP, while all other structural designs remain unchanged. CERP is compared with the following lightweight embedding baselines, which are all model-agnostic:
\begin{itemize}
    \item \textbf{ESAPN} \cite{liu_automated_2020}: An AutoML-based dimension search algorithm that utilizes reinforcement learning (RL) to solve the discrete embedding size selection problem.
    \item \textbf{AutoEmb} \cite{zhao_autoemb_2020}: A differentiable, AutoML-based dimension search algorithm that performs soft selection by using the weighted sum on embedding vectors with different dimension sizes.
    \item \textbf{PEP} \cite{liu_learnable_2021}: An automatic embedding sparsification technique that solely relies on $L_1$ regularized pruning to reach desired memory budget.
    \item \textbf{QR} \cite{shi_compositional_2020}: A work that deploys compositional embedding structure via quotient-remainder hashing trick for meta-embedding indexing.
    \item \textbf{OptEmbed} \cite{lyu_optembed_2022}: The state-of-the-art embedding optimization framework that combines pruning and AutoML-based embedding dimension search.
    \item \textbf{DHE} \cite{kang_learning_2021}: A hashing-based technique that replaces the entire embedding table with fixed-length hash codes and a DNN network to compute unique embedding vectors.
\end{itemize}
Besides, for both base recommenders MLP and LightGCN, we implement a vanilla baseline with a fixed embedding size for all users and items, where the fixed embedding size is set to the maximum integer allowed for each given sparsity rate. We term this baseline the \textbf{unified dimensionality (UD)} approach.

\subsubsection{Evaluation Metrics}
We follow the common practice in recommendation research \cite{he_lightgcn_2020,hung2017computing,chen2021temporal, wang_neural_2019} to use $\textit{NDCG}@N$ \cite{wang2013theoretical} and $Recall@N$ as evaluation metrics and $N$ is fixed to $10$. For UD, PEP, QR and CERP, to testify their effectiveness under different memory budgets, we choose three embedding sparsity rates $s\in \{90\%, 95\%, 99\%\}$, where the compressed embedding table is guaranteed to have no more than $sd(|\mathcal{U}|+|\mathcal{I}|)$ parameters. Notably, ESAPN, AutoEmb, OptEmbed and DHE have a more performance-focused optimization objective and lack a mechanism to precisely control the resulting embedding sparsity, hence we tune them to obtain a sparsity as close to $90\%$ as possible and only report their performance under their final sparsity achieved.

\begin{table*}[t!]
\caption{Performance comparison between default settings and settings with particular component modified. Default means settings with $\mathcal{L}_{prune}$ and exponential decay on $\gamma$ enabled, and the bucket size is balanced. Overlap rate in our context refers to the percentage of overlapping non-zero dimensions between the two meta-embeddings used for composing all entity embeddings.}
\label{tab:rq2_overall}
\centering
\resizebox{\textwidth}{!}{
\setlength\tabcolsep{1.8pt}
\begin{tabular}{cc|cccccc||cccccc}
\hline
 &
   &
  \multicolumn{6}{c||}{Gowalla} &
  \multicolumn{6}{c}{Yelp2020} \\ \cline{3-14} 
 &
   &
  \multicolumn{3}{c|}{MLP} &
  \multicolumn{3}{c||}{LightGCN} &
  \multicolumn{3}{c|}{MLP} &
  \multicolumn{3}{c}{LightGCN} \\ \hline
\multicolumn{1}{c|}{Sparsity} &
  Variant &
  \begin{tabular}[c]{@{}c@{}}Avg\\Dim\end{tabular} &
  N@10 &
  \multicolumn{1}{c|}{\begin{tabular}[c]{@{}c@{}}Overlap\\Rate\end{tabular}} &
  \begin{tabular}[c]{@{}c@{}}Avg\\Dim\end{tabular} &
  N@10 &
  \begin{tabular}[c]{@{}c@{}}Overlap\\Rate\end{tabular} &
  \begin{tabular}[c]{@{}c@{}}Avg\\Dim\end{tabular} &
  N@10 &
  \multicolumn{1}{c|}{\begin{tabular}[c]{@{}c@{}}Overlap\\Rate\end{tabular}} &
  \begin{tabular}[c]{@{}c@{}}Avg\\Dim\end{tabular} &
  N@10 &
  \begin{tabular}[c]{@{}c@{}}Overlap\\Rate\end{tabular} \\ \hline
\multicolumn{1}{c|}{\multirow{4}{*}{90\%}} &
  Default &
  124.33 &
  0.0312 &
  \multicolumn{1}{c|}{49.17\%} &
  125.39 &
  0.0965 &
  47.89\% &
  122.94 &
  0.0061 &
  \multicolumn{1}{c|}{51.80\%} &
  125.79 &
  0.0230 &
  49.85\% \\ \cline{2-14} 
\multicolumn{1}{c|}{} &
  w/o $\mathcal{L}_{prune}$&
  118.23 &
  0.0310 &
  \multicolumn{1}{c|}{53.40\%} &
  119.34 &
  0.0658 &
  52.93\% &
  118.91 &
  0.0049 &
  \multicolumn{1}{c|}{54.62\%} &
  118.70 &
  0.0069 &
  54.93\% \\ \cline{2-14} 
\multicolumn{1}{c|}{} & w/o $\gamma$ decay&
  125.52 &
  0.0302 &
  \multicolumn{1}{c|}{49.07\%} &
  125.93 &
  0.0969 & 47.59\%
   &
  123.07 &
  0.0053 &
  \multicolumn{1}{c|}{47.58\%} &
  126.01 &
  0.0222 & 49.91\%
   \\ \cline{2-14} 
\multicolumn{1}{c|}{} &
  Imbalanced meta-embeddings &
  128 &
  0.0313 &
  \multicolumn{1}{c|}{70.75\%} &
  128 &
  0.0815 &
  70.79\% &
  128 &
  0.0056 &
  \multicolumn{1}{c|}{72.58\%} &
  128 &
  0.0200 &
  72.60\% \\ \hline\hline
\multicolumn{1}{c|}{\multirow{4}{*}{95\%}} &
  Default &
  83.96 &
  0.0307 &
  \multicolumn{1}{c|}{8.07\%} &
  84.88 &
  0.0913 &
  6.80\% &
  80.85 &
  0.0057 &
  \multicolumn{1}{c|}{10.94\%} &
  82.37 &
  0.0206 &
  9.89\% \\ \cline{2-14} 
\multicolumn{1}{c|}{} &
  w/o $\mathcal{L}_{prune}$ &
  74.39 &
  0.0295 &
  \multicolumn{1}{c|}{14.83\%} &
  78.99 &
  0.0665 &
  11.99\% &
  75.80 &
  0.0057 &
  \multicolumn{1}{c|}{14.40\%} &
  73.40 &
  0.0070 &
  16.59\% \\ \cline{2-14} 
\multicolumn{1}{c|}{} & w/o $\gamma$ decay&
  \multicolumn{3}{c|}{\begin{tabular}[c]{@{}c@{}}Sparsity stalls at 94.94\% \end{tabular}} &
  \multicolumn{3}{c||}{Sparsity stalls at 93.86\%} &
  81.18 &
  0.0061 &
  \multicolumn{1}{c|}{4.16\%} &
  82.58 &
  0.0189 & 6.43\%
   \\ \cline{2-14} 
\multicolumn{1}{c|}{} &
  Imbalanced meta-embeddings &
  128 &
  0.0293 &
  \multicolumn{1}{c|}{35.35\%} &
  128 &
  0.0743 &
  35.40\% &
  128 &
  0.0059 &
  \multicolumn{1}{c|}{36.31\%} &
  128 &
  0.0191 &
  36.31\% \\ \hline \hline
\multicolumn{1}{c|}{\multirow{4}{*}{99\%}} &
  Default &
  18.30 &
  0.0298 &
  \multicolumn{1}{c|}{0.43\%} &
  17.99 &
  0.0831 &
  0.76\% &
  18.66 &
  0.0061 &
  \multicolumn{1}{c|}{0.32\%} &
  18.61 &
  0.0167 &
  0.41\% \\ \cline{2-14} 
\multicolumn{1}{c|}{} &
  w/o $\mathcal{L}_{prune}$ &
  17.10 &
  0.0287 &
  \multicolumn{1}{c|}{1.26\%} &
  17.24 &
  0.0646 &
  1.33\% &
  16.27 &
  0.0058 &
  \multicolumn{1}{c|}{2.00\%} &
  15.39 &
  0.0069 &
  2.76\% \\ \cline{2-14} 
\multicolumn{1}{c|}{} & w/o $\gamma$ decay &
  \multicolumn{3}{c|}{\begin{tabular}[c]{@{}c@{}}Sparsity stalls at 94.94\% \end{tabular}} &
  \multicolumn{3}{c||}{\begin{tabular}[c]{@{}c@{}}Sparsity stalls at 93.86\% \end{tabular}} &
  \multicolumn{3}{c|}{\begin{tabular}[c]{@{}c@{}}Sparsity stalls at 97.43\% \end{tabular}} &
  \multicolumn{3}{c}{\begin{tabular}[c]{@{}c@{}}Sparsity stalls at 97.00\% \end{tabular}} \\ \cline{2-14} 
\multicolumn{1}{c|}{} &Imbalanced meta-embeddings &
  128 &
  0.0280 &
  \multicolumn{1}{c|}{7.01\%} &
  127.99 &
  0.0572 &
  7.04\% &
  128 &
  0.0058 &
  \multicolumn{1}{c|}{7.21\%} &
  128 &
  0.0145 &
  7.25\% \\ \hline
\end{tabular}
}
    \vspace*{-4mm}
\end{table*}

\subsubsection{Implementation Details}
 In our work, the full embedding size $d$ for $\mathbf{P}$ and $\mathbf{Q}$ is set to $128$. The bucket size $b$ is set to $5,000$ for Gowalla dataset and $8,000$ for Yelp2020 dataset. All trainable parameters are initialized via Xavier Initialization \cite{glorot2010understanding}. The pruning temperature $\eta$ is fixed at 100. 
 We use Adam optimizer, with the optimal learning rate selected from $\{\expnumber{1}{-2}, \expnumber{1}{-3}, \expnumber{1}{-4}\}$ and weight decay selected from $\{\expnumber{1}{-5}, \expnumber{1}{-6}, \expnumber{1}{-7}\}$. For MLP recommender, the number of hidden layers is set to $3$. The number of message-passing layers in LightGCN is set to $4$. 
For the two search-based methods ESAPN and AutoEmb, we use smaller values in their candidate dimension (\ie action) set, so as to let their final embeddings achieve comparable sparsity rates. For DHE, since it computes embedding vectors by inputting dense hash encodings to DNN, we set the hash code length to 12 to match the lowest sparsity rate of $90\%$. For PEP, we adopt its optimal settings reported in the paper \cite{liu_learnable_2021}. For QR, we adjust the bucket size of the remainder embedding table so that the total parameter size of the two meta-embedding tables can match the budgeted number of parameters. 

\begin{figure}[t]
        \centering
        \begin{subfigure}[b]{.235\textwidth}
            \centering
            \includegraphics[width=\textwidth]{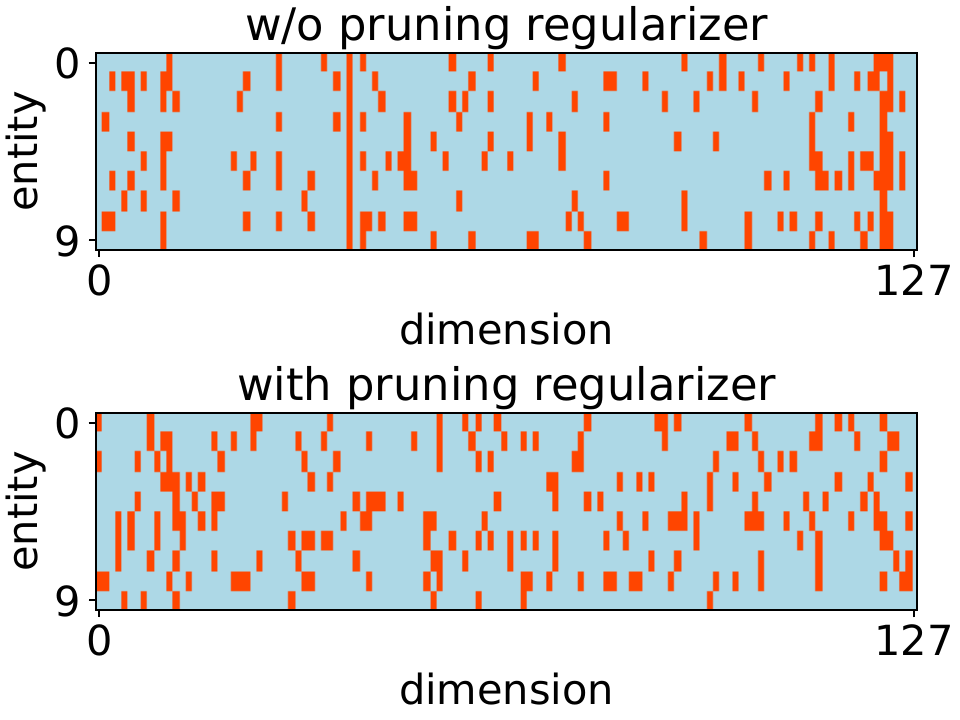}
            \vspace{-0.6cm}
            \caption[]%
            {{\small CERP with MLP}}    
        \end{subfigure}%
        \hspace{0.1cm}
        \begin{subfigure}[b]{.235\textwidth}  
            \centering 
            \includegraphics[width=\textwidth]{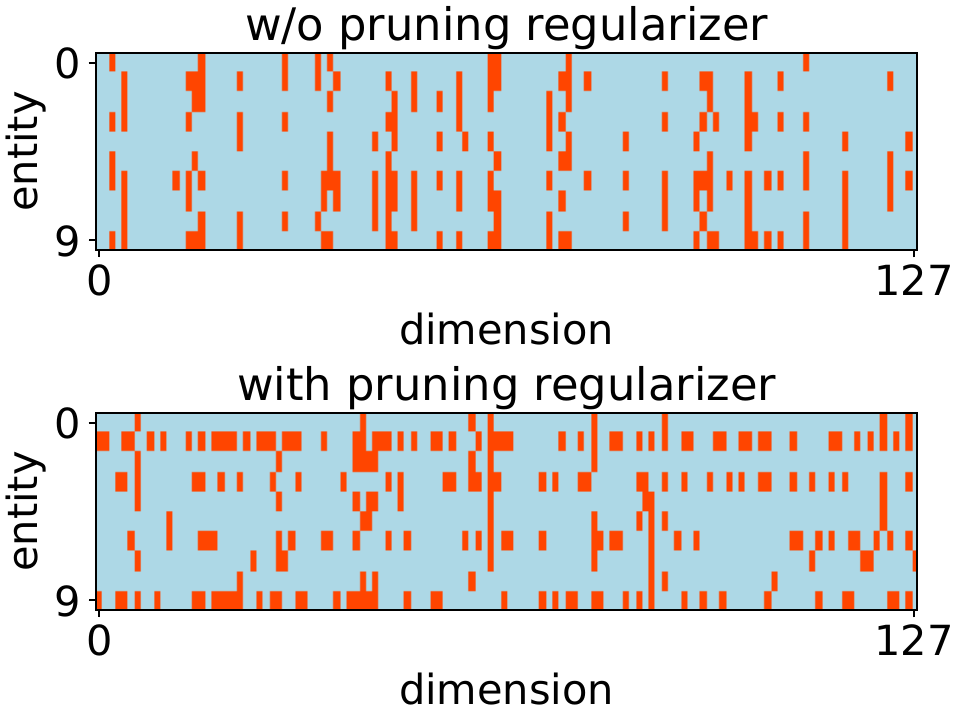}
            \vspace{-0.6cm}
            \caption[]%
            {{\small CERP with LightGCN}}    
        \end{subfigure}%
        \vspace{-2mm}
        \caption[ ]
        {Visualization of sampled entities embedding vectors with and without pruning regularizer in $99\%$ sparsity pruned embedding tables. (*A red cell means a non-zero value in the embedding vector, a light blue cell indicates otherwise.)} 
        \label{fig:visualized_pruning_loss_vectors}
        \vspace*{-6mm}
\end{figure}

\subsection{Overall Performance (RQ1)}
The overall performance benchmark of all tested methods under different memory budgets is shown in Table \ref{tab:overall_performance}. We summarize our findings as follows.

\textbf{Recommendation Performance.} Regardless of the choice of base recommender, our method outperforms all baselines on both datasets. While methods with LightGCN base recommender perform better than those with MLP base recommender across the three memory budgets, LightGCN is more sensitive to the memory budget on fixed size settings or implemented on PEP and QR. Whereas when implemented with CERP, the performance degradation is reduced significantly under a tight memory budget. This can be witnessed as the huge performance percentage increase on both datasets under the $99\%$ sparsity. 
As for methods that cannot precisely control the final embedding size, we find that no methods can obtain comparable results to CERP under the $90\%$ sparsity memory budget. Among them, OptEmbed attains relatively better performance. Both ESAPN's and AutoEmb's performance heavily depends on the choice of the base recommender.
DHE is the weakest method on both datasets, indicating that using dense hash encodings with small dimension sizes does not avoid the expressiveness limitation of generated entity embeddings. 

\textbf{Average Embedding Dimension.} UD, PEP, QR and CERP can meet various memory budgets, while PEP squeezes the average entity embedding size to an extremely low 1.28 at $99\%$ sparsity, compared to a size of around 18 achieved by CERP. As a non-pruning method, QR assigns full size to all embeddings unanimously but still underperforms due to the limited number of meta-embeddings. This demonstrates CERP's strong capability in generating expressive embeddings under an extremely tight memory budget.

\begin{figure*}[t!]
        \centering
        \includegraphics[width=0.85\textwidth]{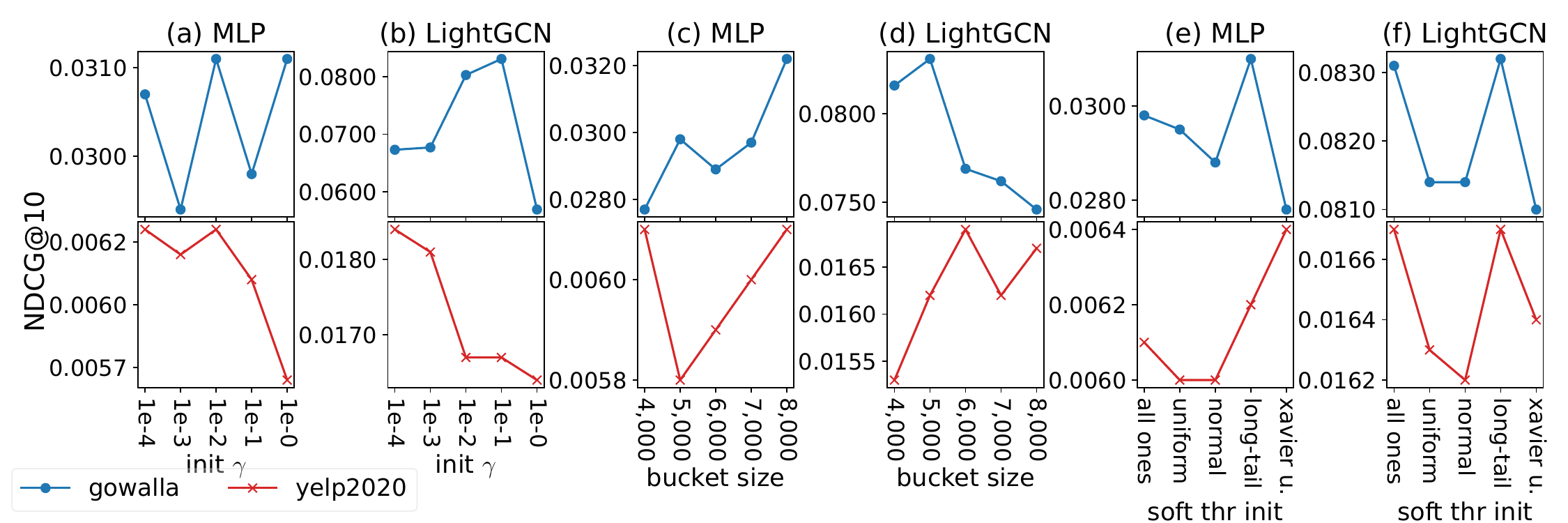}
        \vspace*{-2mm}
        \caption[ ]
        {Performance of CERP w.r.t. different hyperparameter settings and backbone recommenders, where ``soft thr init'' is a shorthand for soft threshold initializer.} 
        \label{fig:hyperparam_performance}  
        \vspace*{-6mm}
\end{figure*}

\subsection{Discussions on Key Model Components (RQ2)}
To validate the performance gain from each key component of CERP, we carry out ablation studies on the pruning regularizer $\mathcal{L}_{prune}$, the exponential decay on $\gamma$, as well as the use of our balanced hashing trick. The performance benchmark used in this section is shown in Figure \ref{tab:rq2_overall}.

\subsubsection{Pruning Regularizer}
To study the impact of pruning regularizer, we conduct pruning on each dataset with and without $\mathcal{L}_{prune}$, and then test their performance. It is discovered that 
the settings with pruning regularizer always obtain a higher average embedding dimension size than those without on both datasets. The model accuracy of settings with the regularizer enabled also improves. 
In terms of non-zero value overlap rates between meta-embedding vectors, applying the pruning regularizer effectively reduces the collision of non-zero dimensions of two meta-embedding vectors, making them complementary. 
As a qualitative analysis, we sample $10$ entities in Gowalla dataset and visualize their composed embeddings from $99\%$ pruned embedding tables in Figure \ref{fig:visualized_pruning_loss_vectors}. It is witnessed that with the regularizer, the pruned embeddings are denser for some particular entities, showing CERP can assign more usable dimensions to important entity embeddings to retain their fidelity.

\subsubsection{Exponential Decay on \texorpdfstring{$\gamma$}{gamma}}
We switch on and off the exponential decay behavior on the control factor of the pruning regularizer $\gamma$ to examine its effects on the quality of pruned embedding table and pruning efficiency. We set a maximum pruning epoch limit of $50$ in case there are settings that never terminate. In terms of pruning efficiency, the exponential decay does not affect the number of pruning epochs required to reach $90\%$ sparsity. However, for more rigorous sparsity targets, settings without exponential decay on $\gamma$ may fail to reach them.
Regarding model accuracy, the $\textit{NDCG}@10$ scores of pruned embedding tables created by settings with exponential decay switched on, in general, are better than those without. Our experiments confirm the necessity of applying exponential decay on pruning regularization loss control factor $\gamma$ so that CERP can switch the goal between preserving embedding quality and accelerating pruning speed.

\subsubsection{Balanced Codebook Hashing}
In CERP, we leverage two codebooks with the same bucket size $b$ for compositional embeddings, such that each meta-embedding is shared by as few entities as possible. Alternatively, Shi \etal \cite{shi_compositional_2020}, suggests an imbalanced bucket size arrangement scheme with a quotient-remainder hashing trick. In their proposal, the bucket size of the quotient embedding table fully depends on the bucket size of the remainder embedding table. Such an arrangement reduces the bucket size of the quotient embedding table significantly.
We implement this bucket size arrangement as well in CERP and test its performance against our balanced setting. 
To make a fair comparison between the two bucket size schemes, we define the bucket size of the remainder embedding table in the imbalanced setting $b' = 2 \times b$. It follows that $b'$ in Gowalla is $10,000$ and in Yelp2020 it is $16,000$. 
One noticeable change with the imbalanced bucket size scheme is that 
the retrieved embedding vectors barely contain zero values despite using the pruning technique. This is due to the fact that the bucket size of the quotient embedding table is only a fraction of the bucket size of the remainder embedding table, which makes the pruning algorithm consider every element in the quotient embedding table to be equally important and hence, should be retained. As shown in Table \ref{tab:rq2_overall}, the following consequence is the degradation in performance. A guess of cause is the high non-zero value collision which lowers the uniqueness of entity embeddings.

\subsection{Hyperparameter Sensitivity (RQ3)}
In this section, we explore our framework CERP's sensitivity to three crucial hyperparameters, namely the initial value of $\gamma$, the bucket size of two codebooks, as well as the initialization method of the soft threshold for pruning. 
We visualize the performance trend 
conducted under the $99\%$ sparsity memory budget in Figure \ref{fig:hyperparam_performance}.

\subsubsection{Initial Value of \texorpdfstring{$\gamma$}{gamma}}
The set of initial $\gamma$ values for testing is $\{\expnumber{1}{-4}, \expnumber{1}{-3}, \expnumber{1}{-2}, \expnumber{1}{-1}, 1\}$. Figure \ref{fig:hyperparam_performance}a and Figure \ref{fig:hyperparam_performance}b show the performance of MLP and LightGCN respectively. 
The MLP settings are generally less sensitive to the initial value of $\gamma$ than LightGCN settings. This is especially true on Gowalla dataset,
indicating the initial value of $\gamma$ is crucial to settings with LightGCN base recommender, especially for datasets with relatively dense entity interactions.

\subsubsection{Bucket Size}
We choose the bucket size from $\{$4,000; 5,000; 6,000; 7,000; 8,000$\}$ for both codebooks to conduct this part of hyperparameter testing. The performance results on both datasets are revealed in Figure \ref{fig:hyperparam_performance}c and Figure \ref{fig:hyperparam_performance}d. 
It is discovered that 
blindly increasing the bucket size does not guarantee performance improvement. This is mainly because under a fixed memory constraint, CERP faces the trade-off between embedding uniqueness and embedding fidelity. The higher the bucket size yields the sparser the meta-embedding vectors for entities, so are their compositional embeddings. Hence, the number of usable parameters is sacrificed. Too few buckets used on the other hand, hurts embedding uniqueness.

\subsubsection{Soft Threshold Initialization}
To study the impact of the soft threshold's initial values on embedding quality, we conduct experiments by setting all values in the soft threshold base matrix to one (termed ``all ones'' initialization) or randomize it using the Uniform distribution, Normal distribution, Long-tail distribution, or Xavier Uniform distribution \cite{glorot2010understanding}. 
The performance trend is depicted in Figure \ref{fig:hyperparam_performance}e and Figure \ref{fig:hyperparam_performance}f. We find that most settings using the long-tail distribution perform exceptionally better than others. In settings with LightGCN base recommender, the simple ``all ones'' initialization is sufficient for attaining comparable results to the setting which uses the long-tail distribution.

\section{Conclusion} \label{sec:conclusion}
In this paper, we acknowledge the embedding table's space efficiency dilemma in latent factor recommender models and identify the possible side effects that arise by devising contemporary embedding optimization techniques. We propose a novel compact embedding framework CERP to overcome recognized challenges. In CERP, we take advantage of the benefits of dynamic embedding size allocation and compositional embeddings. We design an innovative regularizer to enforce complementary behavior between the two types of work. Our comprehensive experiments confirm CERP's capability to obtain embedding tables that satisfy various memory budgets. The performance results also indicate the superiority of CERP over other embedding optimization baselines. 

\section {Acknowledgments}
This work is supported by the Australian Research Council under the streams of Future Fellowship (No. FT210100624), Discovery Project (No. DP190101985), and Discovery Early Career Research Award (No. DE200101465 and No. DE230101033).

\bibliographystyle{IEEEtran}
\bibliography{customized}

\end{document}